\documentclass[11pt,letter]{article}
\pdfoutput=1
\usepackage{graphicx}
\usepackage{epsfig}
\usepackage{amsmath,amssymb,hyperref,enumerate}
\usepackage{slashed}
\usepackage{amsfonts}
\usepackage{url}
\usepackage{color}
\usepackage{appendix}
\usepackage{bm}
\usepackage{subfig}
\usepackage{cite}
\hypersetup{colorlinks,citecolor= black,linkcolor= black}
\definecolor{nicered}{rgb}{0.7,0.1,0.1}
\definecolor{nicegreen}{rgb}{0.1,0.5,0.1}

\usepackage[normalem]{ulem}
\usepackage{setspace}

\def\({\left(}
\def\){\right)}
\def\[{\left[}
\def\]{\right]}

\usepackage[margin=1.in]{geometry}

\allowdisplaybreaks

\begin{document}

\singlespacing
\onehalfspacing

\renewcommand{\thefootnote}{\fnsymbol{footnote}}

\thispagestyle{empty}

\begin{flushright}
OSU-HEP-13-07
\end{flushright}

\vspace*{2.0cm}
\begin{center}
\LARGE {\bf \Large Predictive Model of Radiative Neutrino Masses}
\end{center}

\begin{center}
{\large \bf K.S. Babu$^a$\footnote{Email:
babu@okstate.edu} and J. Julio$^b$\footnote{Email:
julio@ijs.si}}
\end{center}

\vspace*{-0.2in}
\begin{center}
\it $^a $Department of Physics, Oklahoma State University, Stillwater, Oklahoma 74078, USA

\vspace*{-0.1in}
\bigskip
\it $^b $Jo\v zef Stefan Institute, Jamova cesta 39, Ljubljana 1001, Slovenia
\end{center}

\renewcommand{\thefootnote}{\arabic{footnote}}
\setcounter{footnote}{0}

\begin{abstract}
We present a simple and predictive model of radiative neutrino masses. It is a special case of the Zee model which introduces two Higgs doublets and a charged singlet. We impose a family-dependent $Z_4$ symmetry acting on the leptons, which reduces the number of parameters describing neutrino oscillations to four. A variety of predictions follow: The hierarchy of neutrino masses must be inverted; the lightest neutrino mass is extremely small and calculable; one of the neutrino mixing angles is determined in terms of the other two; the phase parameters take CP--conserving values with $\delta_{CP} = \pi$; and the effective mass in neutrinoless double beta decay lies in a narrow range, $m_{\beta \beta} = (17.6 - 18.5)$ meV. The ratio of vacuum expectation values of the two Higgs doublets, $\tan\beta$, is determined to be either $1.9$ or $0.19$ from neutrino oscillation data.  Flavor-conserving and flavor-changing couplings of the Higgs doublets are also determined from neutrino data.  The non-standard neutral Higgs bosons, if they are moderately heavy, decay dominantly into $\mu$ and $\tau$ with prescribed branching ratios.  Observable rates for the decays  $\mu \rightarrow e \gamma$ and $\tau \rightarrow 3\mu$ are predicted if these scalars have masses in the range of $150-500$ GeV.

\end{abstract}

\newpage

\section{Introduction}

The purpose of this paper is present a simple model of radiative neutrino masses.  The model is a special
case of the Zee model \cite{Zee:1980ai}.  It assumes the existence of two Higgs doublets and a charged singlet.  When two Higgs doublets are present
in the Standard Model, in general there are Higgs mediated flavor changing neutral currents (FCNC) at the tree--level.  In the original
Zee model such FCNC were allowed.  While the model provides a simple way for explaining small neutrino masses with TeV scale physics, in this general setup
testing the model quantitatively becomes difficult.  Soon after the Zee model was proposed, Wolfenstein suggested \cite{Wolfenstein:1980sy}
that there is a discrete $Z_2$ symmetry in the model which would forbid tree--level FCNC mediated by the Higgs bosons \cite{Glashow:1976nt}.
The Zee-Wolfenstein model is quite predictive in the neutrino sector, and was very popular for a long time \cite{frampton}.  After more precise solar neutrino and KAMLAND data emerged, it became clear that the Zee-Wolfenstein model cannot support the oscillation data \cite{He:2003ih}.  The chief reason for the exclusion was a special feature this model has, namely the diagonal elements of the neutrino mass matrix all vanish in the flavor basis. Attention has moved on to other
interesting models of radiative neutrino mass generation, especially since these models may be testable at the LHC as well as in lepton flavor violation processes \cite{alternative1,alternative2}.

Perhaps the $Z_2$ symmetry assumed in the Zee model is too strong. With no additional symmetries however, the model is not so predictive, and it is not clear
how to test it quantitatively.  The model we present here is  a specific realization of the Zee model which is in between the two
extremes of having no symmetry at all and having no tree-level FCNC at all.  We assume a discrete symmetry in the model, but unlike Wolfenstein, we allow it to be family-dependent.  This would indeed lead to tree-level FCNC mediated by the Higgs bosons, but the amplitudes for such processes are sufficiently small and consistent with experimental
constraints, even when the Higgs bosons have masses of about 100 GeV.

Our model is the Zee model with a $Z_4$ symmetry acting on the leptons and the Higgs bosons.  In the quark sector one Higgs doublet
couples universally to the up--type and down--type quarks.  With this assignment the $Z_4$ symmetry is anomaly free \cite{discrete}.
As a result of the
structure of the model and the $Z_4$ symmetry, all of neutrino oscillation data is described in terms of four real parameters.  There are
then a variety of predictions.  Neutrino mass hierarchy is predicted to be inverted.  The CP violation parameter is predicted to be
$\delta_{CP} = \pi$.
Among the three neutrino oscillation angles, one is determined in terms of the other two.  This relation is $|U_{\tau1}| = |U_{\tau 2}|$, which is found to work well with present data. The effective mass for neutrinoless double beta decay lies in a narrow range, $m_{\beta \beta} = (17.6 - 18.5)$ meV.

A fit to the neutrino oscillation data also determines the parameter $\tan\beta$, the ratio of the two neutral Higgs vacuum expectation values.
We find two solutions, $\tan\beta = 0.19$ or $1.9$.  This enables us to calculate the branching ratios of the moderately heavy Higgs bosons
decaying into fermions.  Leptonic decay modes are significant, especially with muons in the final state.  Flavor violation mediated by
the Higgs bosons in the lepton sector is also calculable.  Rates for the decays $\mu \rightarrow e\gamma$ and $\tau \rightarrow 3\mu$ may
be accessible to proposed experiments.

The rest of the paper is organized as follows.  In Sec. 2 we describe the model.  In Sec. 3 we address the flavor structure of the charged
lepton mass matrix and Yukawa matrices.  Neutrino phenomenology is worked out in Sec. 4.  Sec. 5 is devoted to lepton flavor violation
discussions, and Sec. 6 addresses Higgs decays.  Finally, Sec. 7 has our conclusions.

\section{The Model}

The model we present is a special case of the general Zee model \cite{Zee:1980ai}.  Neutrino masses are induced
as one--loop radiative corrections through the exchange of charged scalars.  The gauge symmetry
and the fermionic content of the model are identical to that of the Standard Model. In particular,
Standard Model singlet right--handed neutrinos are not introduced.
The scalar sector is extended so that there are two Higgs doublets $H_a(1,2,-1/2)$ ($a=1,2$)
and a charged singlet $\eta^+(1,1,+1)$.  A discrete $Z_4$ symmetry acting on the leptons fields $L_i(1,2,-1/2)$,
$e^c_i(1,1,+1$) and the
Higgs fields $H_a$ and $\eta^+$ is assumed, with the following transformation properties:
\begin{eqnarray}
&~& L_i: (-i,\, i,\, i);~~~~~e^c_i: (-i,\, -i,\, -i); \nonumber\\
&~& H_1: +1;~~~H_2: -1;~~~\eta^+: -1~.
\label{Z4charge}
\end{eqnarray}
Here $i=1-3$ is the family index.  Thus the $Z_4$ symmetry is family-dependent.  This is the crucial
difference of our model compared to the Wolfenstein realization of the Zee model, where a family
universal $Z_2$ is assumed in order to suppress naturally tree--level flavor changing neutral currents mediated by
the Higgs bosons.  In our version, there will be tree-level flavor changing neutral currents, but as we show,
the amplitudes for these processes are sufficiently suppressed to be consistent with data, even when the neutral
scalars which mediate them have masses of order hundred GeV.

In the leptonic sector the following Yukawa couplings can be written down consistent with the gauge symmetry
and the $Z_4$ symmetry of Eq. (\ref{Z4charge}).
\begin{equation}
{\cal L}_{\rm Yuk}^{(\ell)} = \sum_{\substack{
      i=2,3, \\
      \alpha=1,2,3}} Y_{i\alpha} L_i e^c_\alpha H_1 + \sum_{\alpha=1,2,3} Y_\alpha L_1 e^c_\alpha H_2 + f_{23} L_2 L_3 \eta^+ + h.c.
      \label{Yuk}
\end{equation}
Lepton number is not broken by these Yukawa couplings, as can be seen by assigning lepton number of $-2$ to $\eta^+$ field.
However, the Higgs potential contains a cubic term which is $Z_4$--invariant that breaks lepton number, and possibly
also a quadratic term that breaks the $Z_4$ symmetry softly:
\begin{equation}
V = \left\{\mu H_1 H_2 \eta^+ + m_{12}^2 H_1^\dagger H_2 + h.c.\right\} + ....
\label{V0}
\end{equation}
Here the $....$ stands for other terms which are not so relevant for our present discussions.  However, it should
be noted that the action of the $Z_4$ symmetry does not create an accidental global $U(1)$ symmetry of the Higgs potential,
which could have led to an unwanted pseudo-Goldstone boson.  (Note that the $Z_4$ symmetry allows a quartic coupling $(H_1^\dagger H_2)^2 + h.c$
in the Higgs potential which guarantees that there is no global $U(1)$ present, even in the absence of soft breaking of $Z_4$ by
the $m_{12}^2$ term of Eq. (\ref{V0}).) In our discussions we shall allow for
$m_{12}^2$ in Eq. (\ref{V0}) to be either zero or nonzero, keeping the option open for breaking the $Z_4$ symmetry softly.
The two cases lead to essentially the same results in the neutrino sector, but would affect the Higgs phenomenology differently.

In the quark sector the $Z_4$ symmetry of the model acts universally with all the down-type quarks
and the up-type quarks coupling to the same Higgs field $H_1$ or $H_2$.  The quark Yukawa couplings
have the form
\begin{equation}
{\cal L}^{(q)}_{\rm Yuk} = \sum_{i,j=1-3} Y_{ij}^u Q_i u^c_j \tilde{H}_a +\sum_{i,j=1-3}Y_{ij}^d Q_i d^c_j H_a + h.c.
\label{quark}
\end{equation}
where the Higgs label $a$ takes the same value, either 1 or 2, in both terms.  Here $\tilde{H}_a = i \tau_2 H_a^*$.  With this form of the
quark Yukawa couplings the $Z_4$ charge assignment of Eq. (\ref{Z4charge}) is anomaly-free \cite{discrete}.
To see this, consider the case where the Higgs field $H_a$
in both terms of Eq. (\ref{quark}) is $H_1$.  In this case, the following $Z_4$ charges can be
assigned to the quarks:  $Q_i: (-i,\,-i,\,-i)$, $u^c_i: (i,\,i,\,i)$, and $d^c_i: (i,\,i,\,i)$.  The mixed
$[SU(3)]^2 \times Z_4$ and $[SU(2)_L]^2 \times Z_4$ anomaly coefficients are then
\begin{eqnarray}
A_2[(SU(2)_L)^2 \times Z_4] &=& \frac{1}{2} \left\{(-1+1+1) +3 (-1-1-1) \right\} = -4, \label{weak}\\
A_3[(SU(3)_C)^2 \times Z_4] &=& \frac{1}{2} \left\{2(-1-1-1) +(1+1+1) + (1+1+1)\right\} = 0. \label{color}
\end{eqnarray}
In Eqs. (\ref{weak})-(\ref{color}), the factor $\frac{1}{2}$ is the index of the fundamental representation of $SU(N)$, the factors 3 and 2 are color and $SU(2)_L$ multiplicities, and a $Z_4$ charge of $\pm i$ is treated as charge $\pm 1$ mod(4).  Now, the condition for the absence of discrete anomalies for a $Z_N$ group is that all the anomaly coefficients must obey $A_i = p_i (N/2)$ with $p_i$ being  integers.  We see that both anomalies satisfy this condition.  The $[U(1)_Y]^2 \times Z_4$ anomaly coefficient is not restricted by the discrete anomaly cancelation condition.
If all quarks couple to $H_2$ in Eq. (\ref{quark}) instead of $H_1$, the $Z_4$ charge assignment of $Q_i: (-i,\,-i,\,-i)$, $u^c_i: (-i,\,-i,\,-i)$, and $d^c_i: (-i,\,-i,\,-i)$
can be chosen, in which case Eq. (\ref{weak}) will remain unchanged, while Eq. (\ref{color}) will be modified to
\begin{equation}
A_3[(SU(3)_C)^2 \times Z_4] = \frac{1}{2} \left\{2(-1-1-1) +(-1-1-1) + (-1-1-1)\right\} = -6.
\end{equation}
This choice also satisfies $A_3 = p_3/2$, with $p_3=-12$, showing the cancelation of the $Z_4$ anomalies. It is interesting to note that  the
case where the up and down type  quarks couple to different $H_a$ fields universally is incompatible with a non-anomalous $Z_4$ symmetry.

\section{Leptonic mass matrix and Yukawa couplings}

In this section we carry out the diagonalization of the charged leptonic mass matrix and then evaluate the Yukawa coupling matrices in a basis where charged lepton mass matrix is diagonal. The $2 \times 3$ Yukawa coupling matrix with elements $Y_{i\alpha}$ of Eq. (\ref{Yuk}) can be brought to a diagonal form with only $Y_{22}$ and $Y_{33}$ elements being nonzero by flavor rotations among the $L_{2,3}$ fields and the $e^c_{1,2,3}$ fields.  The vector with elements $Y_\alpha$ of the second term in Eq. (\ref{Yuk}) will maintain its form with redefined elements.  The form of the $\eta^+$
coupling in Eq. (\ref{Yuk}) will not change by this rotation.  We shall work in a basis where such rotations have been done.

We denote the vacuum expectation values of the neutral components of $H_1$ and $H_2$ as
\begin{equation}
\left\langle H_1^0 \right \rangle = v_1,~~ \left\langle H_2^0 \right \rangle = v_2 =  |v_2| \,e^{i \phi}~
\end{equation}
where $v_1$ has been made real by an $SU(2)_L \times U(1)_Y$ gauge rotation.
Without the soft breaking of $Z_4$  via the $m_{12}^2$ term in Eq. (\ref{V0}), the phase of $v_2$ will
be minimized to $\phi = 0 ~{\rm or}~  \pi$.  With non-vanishing $m_{12}^2$, a nonzero $\phi$ will result, which is determined in terms of the
Higgs potential parameters.  We shall assume a generic nonzero $\phi$, but note that the special case where $m_{12}^2=0$ can be recovered by
setting $\phi=0$ or $\pi$.

The charged lepton mass matrix that results from Eq. (\ref{Yuk}) can now be written down, in the notation ${\cal L} \supset\ell_i \, (M_\ell)_{ij} \,e^c_j$, as
\begin{eqnarray}
M_{\ell} = \left( \begin{matrix}
m_e^0 \sqrt{1+|x|^2+|y|^2} & ~~~m_\mu^0  \frac{y \sqrt{1+|x|^2}} {\sqrt{1+|x|^2+|y|^2}} & ~~~m_\tau^0 \frac{x}{\sqrt{1+|x|^2}}
\\[0.15in] 0 & ~~~m_\mu^0 \frac{\sqrt{1+|x|^2}}{\sqrt{1+|x|^2+|y|^2}} &~~~ 0 \\[0.15in]~~~ 0~~~ & ~~~~~0~~~ & m_\tau^0 \frac{1}{\sqrt{1+|x|^2}}
\end{matrix}\right)~.
\label{Mlep}
\end{eqnarray}
Here we have defined the parameters $m_e^0,\,m_\mu^0,\, m_\tau^0$ and  $x,\,y$ as
\begin{eqnarray}
&~& Y_1 v_2 = m_e^0 \sqrt{1+|x|^2+|y|^2}, ~~ Y_2 v_2 = m_\mu^0  \frac{y\,\sqrt{1+|x|^2}}{\sqrt{1+|x|^2+|y|^2}},~~Y_3 v_2 = m_\tau^0
\frac{x}{\sqrt{1+|x|^2}},\nonumber \\
&~& Y_{22} v_1 = m_\mu^0 \frac{\sqrt{1+|x|^2}}{\sqrt{1+|x|^2+|y|^2}},~~
Y_{33} v_1 = \frac{m_\tau^0}{\sqrt{1+|x|^2}}.
\end{eqnarray}
$(m_e^0,\,m_\mu^0,\,m_\tau^0)$ will turn out to be the approximate eigenvalues, to be identified as  $(m_e,\,m_\mu,\,m_\tau)$.
Indeed, in the approximation $m_e^0 \ll m_\mu^0 \ll m_\tau^0$, and with the
restriction $x,y \ll m_\mu/m_e \simeq 205$ (so that the (1,1) element of $M_\ell$ does not contribute significantly
to  $m_\mu$), the eigenvalues of $M_\ell$ are simply $m_e = m_e^0,\,m_\mu = m_\mu^0,\,
m_\tau = m_\tau^0$, up to relative corrections of order $(m_e/m_\mu)^2$ for $m_e$ and $(m_\mu/m_\tau)^2$ in $m_\mu$ and $m_\tau$.
Note that in Eq. (\ref{Mlep}) we have allowed the off-diagonal entries to be as large as they can be, consistent with the
requirement of hierarchical eigenvalues, $m_e \ll m_\mu \ll m_\tau$.  The parameters
$x$ and $y$ are allowed to take values of order one, or even larger, subject to $x,\,y \ll m_\mu/m_e \simeq 205$.

In diagonalizing $M_\ell$, we first make phase rotations on the lepton fields to make $x,\,y,\,m_e^0$,
$m_\mu^0,\,m_\tau^0$ all
real.  In this case the square root factor $\sqrt{1+|x|^2}$ can be replaced by $\sqrt{1+x^2}$ and similarly $\sqrt{1+|x|^2+|y|^2}$ by
$\sqrt{1+x^2+y^2}$.  With this phase convention, $M_\ell$ can be diagonalized as
\begin{equation}
U_L^T M_\ell U_R = {\rm diag}.(m_e^0,~m_\mu^0,~m_\tau^0) + {\cal O}\left(\frac{m_i^2}{m_j}\right)  \equiv M_\ell^{\rm diag} = {\rm diag}.(m_e,~m_\mu,~m_\tau),
\end{equation}
where
\begin{eqnarray}
U_L = \left(\begin{matrix} \frac{1}{\sqrt{1+x^2+y^2}}~~~ & \frac{y}{\sqrt{1+x^2} \sqrt{1+x^2+y^2}} & ~~~ \frac{x}{\sqrt{1+x^2}} \cr
\frac{-y}{\sqrt{1+x^2+y^2}}~~~ & \frac{\sqrt{1+x^2}}{\sqrt{1+x^2+y^2}} & ~~~0 \cr
\frac{-x}{\sqrt{1+x^2+y^2}}~~~ & \frac{-x y}{\sqrt{1+x^2}\sqrt{1+x^2+y^2}} & ~~~ \frac{1}{\sqrt{1+x^2}} \end{matrix}\right) +
{\cal O}\left(\frac{m_i^2}{m_j^2}\right),
\end{eqnarray}
and
\begin{eqnarray}
U_R = \left(\begin{matrix} 1 ~~~ & \frac{m_e}{m_\mu}\frac{y}{\sqrt{1+x^2}} & ~~~ \frac{m_e}{m_\tau} \frac{x \sqrt{1+x^2+y^2}}{\sqrt{1+x^2}} \cr
-\frac{m_e}{m_\mu} \frac{y}{\sqrt{1+x^2}}~~~ & 1 & ~~~\frac{m_\mu}{m_\tau} \frac{xy}{\sqrt{1+x^2+y^2}} \cr
-\frac{m_e}{m_\tau} \frac{x \sqrt{1+x^2}}{\sqrt{1+x^2+y^2}}~~~ & -\frac{m_\mu}{m_\tau} \frac{xy}{\sqrt{1+x^2+y^2}} & ~~~ 1 \end{matrix}\right) +
{\cal O}\left(\frac{m_i^2}{m_j^2}\right).
\end{eqnarray}
Here terms that are dropped in $U_L$ and $U_R$ are of order $(m_e^2/m_\mu^2)$ and $(m_\mu^2/m_\tau^2)$.
The matrix $U_L$ is a product of two rotation matrices.
The first rotation by and angle $\theta$ with $\tan\theta = x$ brings the third column of $M_\ell$ to a form $(0,0,m_\tau)^T$.  The second
rotation is by an angle that removes the (rotated) $(1,2)$ entry of $M_\ell$.  The matrix $U_R$ is obtained by sequential rotations in the
$(2-3)$, $(1-2)$ and $(1-3)$ sectors respectively.

Making the same rotation on the matrix which follows from the $L_2L_3$ coupling of the $\eta^+$ field of Eq. (\ref{Yuk}), we obtain the redefined flavor-antisymmetric matrix, $\hat{f}= U_L^T f U_L$, written in the mass eigenbasis for the charged leptons with its elements given by
\begin{eqnarray}
\hat{f} = f_{23} \left( \begin{matrix} 0 & \frac{x}{\sqrt{1+x^2}} & -\frac{y}{\sqrt{1+x^2}\sqrt{1+x^2+y^2}} \cr
-\frac{x}{\sqrt{1+x^2}} & 0 & \frac{1}{\sqrt{1+x^2+y^2}} \cr
\frac{y}{\sqrt{1+x^2}\sqrt{1+x^2+y^2}} & -\frac{1}{\sqrt{1+x^2+y^2}} & 0   \end{matrix} \right).
\label{fhat}
\end{eqnarray}
Note that there is no particular hierarchy factor of the type $(m_e/m_\mu)$ or $(m_\mu/m_\tau)$ that appears in any of the elements of $\hat{f}$.
This feature is central to providing a successful description of neutrino oscillations, as shown in the next section.  Alternative
identifications of mass hierarchies in $M_\ell$, such as the first row entries all being of order $m_\tau$ along with the (2,2) and (3,3) entries
being of order $m_\mu$ and $m_e$, will not preserve this feature, and will be disfavored by neutrino oscillation data.

Among the charged scalar fields $H_1^+$ and $H_2^+$, one combination $G^+=(v_1 H_1^+ + v_2 H_2^+)/v$ (where
$v \equiv \sqrt{v_1^2+|v_2|^2}$) is the Goldstone
boson eaten up by the $W^+$ gauge boson. The orthogonal combination  $H^+ = (v_2^* H_1^+ - v_1 H_2^+)/v$
is physical, which however mixes with the $\eta^+$ field through the cubic scalar coupling of Eq. (\ref{V0}).  The couplings of
the $H^\pm$ fields with leptons can be obtained in the unitary gauge by setting $G^\pm = 0$.  Then we have $H_1^+ = (v_2/v) H^+$
and $H_2^+ = -(v_1/v) H^+$.  The Yukawa couplings of $H^\pm$ before any rotations are done on the lepton fields, except for the phase
rotations that brought $M_\ell$ to a real matrix,  has the form
$\nu^T\, Y_{\rm Yuk}^{(H^\pm)}\, \ell^c H^- + h.c.$ where
\begin{eqnarray}
Y_{\rm Yuk}^{(H^\pm)} = \frac{1}{v} \left( \begin{matrix} -\frac{v_1}{v_2} & ~ & ~ \cr ~ & \frac{v_2^*}{v_1} & ~ \cr
~ & ~ & \frac{v_2^*}{v_1}   \end{matrix} \right) M_\ell~.
\label{charged}
\end{eqnarray}
Note that the diagonal matrix that multiples $M_\ell$ from the left in Eq. (\ref{charged}) only has an overall phase, equal to
$e^{-i\phi}$ where $v_2 = |v_2| e^{i \phi}$.  This overall phase can be absorbed into the definition of $H^-$ field, which
would then make $Y_{\rm Yuk}^{(H^\pm)}$ a real matrix.  In the mass eigenbasis of the charged leptons,
these couplings will become $\hat{Y} = U_L^T\, Y_{\rm Yuk}^{(H^\pm)}\, U_R$.
The elements of $\hat{Y}$ are readily obtained. To leading order in the charged lepton mass ratios ($m_i/m_j$) they are:
\begin{eqnarray}
\hat{Y} = \left[
\begin{matrix}
\frac{m_e}{v} \,\frac{(x^2+y^2) \tan\beta - \cot\beta}{1+x^2+y^2} &
 -\frac{m_\mu}{v} \, \frac{y\,(\tan\beta+\cot\beta)}{\sqrt{1+x^2}\, (1+x^2+y^2)}, &
-\frac{m_\tau}{v} \,\frac{x\,(\tan\beta + \cot\beta)}{\sqrt{1+x^2}\,\sqrt{1+x^2+y^2}} \cr
-\frac{m_e}{v}\, \frac{y\,(\tan\beta + \cot\beta)}{\sqrt{1+x^2}\,(1+x^2+y^2)} &
\frac{m_\mu}{v} \,\frac{\tan\beta\,\{(1+x^2)^2+x^2y^2\}-\cot\beta\, y^2}{(1+x^2)(1+x^2+y^2)} &
-\frac{m_\tau}{v} \, \frac{xy\,(\tan\beta+ \cot\beta)}{(1+x^2)\sqrt{1+x^2+y^2}} \cr
 -\frac{m_e}{v} \, \frac{x\,(\tan\beta+\cot\beta)}{\sqrt{1+x^2}\,\sqrt{1+x^2+y^2}} & -\frac{m_\mu}{v}\,
 \frac{xy\,(\tan\beta + \cot\beta)}{(1+x^2)\,\sqrt{1+x^2+y^2}} &  \frac{m_\tau}{v}\, \frac{\tan\beta - x^2 \cot\beta}{1+x^2}
 \end{matrix} \right].
 \label{Yhat}
\end{eqnarray}
Terms of order $(m_e/m_\mu)^2$ and $(m_\mu/m_\tau)^2$ have been dropped here.  We have defined $\tan\beta \equiv |v_2|/v_1$.

The couplings of the neutral scalar bosons to the lepton fields can be obtained in an analogous way.  If the phase of $v_2$
is nonzero, the physical pseudoscalar boson will mix with the two scalar bosons contained in $H_1^0$ and $H_2^0$.
For real $v_2$ (realized when there is no soft breaking of the $Z_4$ symmetry), such mixings are absent, and
the pseudoscalar Higgs boson $A^0 = [v_2 \,{\rm Im} (H_1^0) - v_1 \,{\rm Im} (H_2^0)]/v$ would couple to the physical leptons
as $\ell_i \hat{Y}_{ij} e^c_j\,A^0/\sqrt{2}$, where $\hat{Y}$ is the same Yukawa coupling matrix as in Eq. (\ref{Yhat}).
And similarly the real scalar boson $H^0 = [v_2 \,{\rm Re} (H_1^0) - v_1 \,{\rm Re} (H_2^0)]/v$ will couple to the physical leptons with the same
Yukawa matrix $\hat{Y}$.  The other (lighter) neutral scalar Higgs boson, $h^0 = [v_1 \,{\rm Re} (H_1^0) + v_2 \,{\rm Re} (H_2^0)]/v$,
is to be identified as the 126 GeV boson discovered at the LHC.  Ignoring $h^0-H^0$ mixing, the state $h^0$ will have
only flavor--diagonal couplings to the leptons, as in the Standard Model.

From Eq. (\ref{Yhat}) and the discussion above, it is clear that the
process $\tau \rightarrow 3 \mu$, mediated by $A^0$ (or $H^0$) would have an amplitude
of order $(m_\mu m_\tau/v^2)/m_{A^0}^2$, which would suppress this process to below the present experimental limit for $m_{A^0}$ of order
a few hundred GeV.  We shall turn to lepton flavor violation in more detail in Sec. 5, after discussing neutrino oscillations,
which would determine the Yukawa matrix $\hat{Y}$ completely.

\section{Neutrino mass generation and phenomenology}

\begin{figure}[t]
\centering
\includegraphics[scale=0.75]{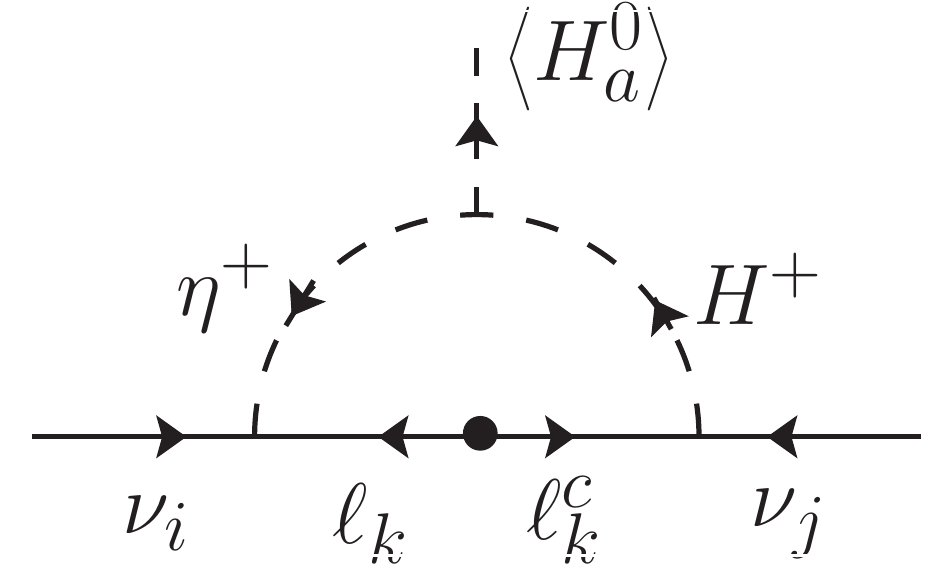}
\caption{One--loop diagram generating neutrino masses in our realization of the Zee model.}
\label{nu-mass-diagram}
\end{figure}

Neutrino masses are generated at the one--loop level by the exchange of charged scalars through diagrams shown in Fig. \ref{nu-mass-diagram}.
There is a second diagram obtained from the diagram shown by replacing the internal particles by their antiparticles.
The mixing of $\eta^+$ and $H^+$ occurs through the cubic
\noindent
scalar coupling in the Higgs potential, see Eq. (\ref{V0}).  We denote the
$\eta^+-H^+$ mixing angle as $\gamma$ and the masses of the physical charged scalar states as $M_1$ and $M_2$.  The induced neutrino
mass matrix is then obtained to be
\begin{equation}
M_\nu = \kappa \left(\hat{f} M_\ell^{\rm diag} \hat{Y}^T + \hat{Y} M_\ell^{\rm diag} \hat{f}^T\right)~.
\label{Mnu}
\end{equation}
Here $M_\ell^{\rm diag} \equiv {\rm diag}. (m_e,\,m_\mu,\,m_\tau)$
is the diagonal charged lepton mass matrix, and $\hat{f}$ and $\hat{Y}$ are the Yukawa coupling
matrices given in Eqs. (\ref{fhat}) and (\ref{Yhat}).  The overall factor $\kappa$ involves the loop integral, and is given by
\begin{equation}
\kappa = \frac{\sin2\gamma }{16 \pi^2}\, {\rm log}\left(\frac{M_1^2}{M_2^2}\right)~.
\end{equation}

The main difference of our realization of the Zee model compared to the Wolfenstein realization \cite{Wolfenstein:1980sy} is the flavor structure of $M_\nu$.
In the Zee--Wolfenstein model the $\hat{Y}$ of Eq. (\ref{Mnu}) is replaced by $M_\ell^{\rm diag}/v$, in which case all diagonal
entries of the neutrino mass matrix would be zero.  Such a mass matrix is now excluded by neutrino oscillation data. In our version, since
$\hat{Y}$ has off-diagonal elements, this will not be the case.  In the general Zee model, $\hat{Y}$ would be a generic matrix,
as opposed to the specific matrix $\hat{Y}$ in Eq. (\ref{Yhat}) here. Thus, in our model, all neutrino data would be determined
by only four parameters: an overall factor $(\kappa f_{23})$, and three parameters $(x,\,y,\,\tan\beta)$ that appear in $\hat{f}$
and $\hat{Y}$.  We now proceed to analyze the predictions of this model for neutrino oscillation parameters.

An interesting feature of $M_\nu$ of Eq. (\ref{Mnu}) is that ${\rm Tr}\,[M_\nu] = 0$.  This can be seen as follows:
\begin{eqnarray}
{\rm Tr}\,[M_\nu] &=& 2\,{\rm Tr}\,[U_L^T f U_L M_\ell^{\rm diag} U_R^T Y_{\rm Yuk}^{H^\pm T} U_L] \nonumber \\
&=& (2 \tan\beta/v)\,{\rm Tr}\,[ f U_L M_\ell^{\rm diag} U_R^T M_\ell^T (1-P)] \nonumber \\
&=& (2 \tan\beta/v)\,{\rm Tr}\,[ f U_L (M_\ell^{\rm diag})^2
U_L^T(1-P)] \nonumber \\
&=& 0\,.
\label{tr}
\end{eqnarray}
Here we defined a diagonal matrix $P = (1+\cot^2\beta)\, {\rm diag}\,(1,\,0,\,0)$.  The last step of Eq. (\ref{tr}) follows by noting that
the first term ${\rm Tr}\,[ f U_L (M_\ell^{\rm diag})^2 U_L^T] = {\rm Tr}\,[\hat{f} (M_\ell^{\rm diag})^2]$ vanishes owing to $\hat{f}$
being antisymmetric.  The second term ${\rm Tr}\,[ f U_L (M_\ell^{\rm diag})^2 U_L^TP]$ also vanishes, since $P$ is nonzero
only in the entry $P_{11}$, while $f$ is nonzero only in the $f_{23}$ and $f_{32}$ entries.  Traceless neutrino mass
matrices have been studied in Ref. \cite{traceless}.

The leading terms of $\hat{Y}$ are in the third column, which all have a factor of $m_\tau$, while the second and third columns, which
have factors of $m_\mu$ and $m_e$ respectively are suppressed. There is no such hierarchical structure in $\hat{f}$.  Noting these
features, we can write down an approximate form for $M_\nu$:
\begin{eqnarray}
(M_\nu)_{ij} \simeq \kappa \left[(a_i b_j + a_j b_i) + (c_i d_j + c_j d_i) \right]
\label{Mnu-approx}
\end{eqnarray}
where
\begin{eqnarray}
\vec{a} = m_\tau \left(\begin{matrix} \hat{f}_{13} \cr \hat{f}_{23} \cr 0 \end{matrix}\right),~~
\vec{b} = \left(\begin{matrix} \hat{Y}_{13} \cr \hat{Y}_{23}  \cr \hat{Y}_{33} \end{matrix}\right),~~
\vec{c} = m_\mu \left(\begin{matrix} \hat{f}_{12} \cr 0  \cr \hat{f}_{32} \end{matrix}\right),~~
\vec{d} = \left(\begin{matrix} \hat{Y}_{12} \cr \hat{Y}_{22}  \cr \hat{Y}_{32} \end{matrix}\right).
\end{eqnarray}
Here we have ignored the contributions proportional to $m_e^2$, which are extremely small.  The terms
$(a_i b_j + a_j b_i)$ in Eq. (\ref{Mnu-approx}) are dominant over the terms $(c_i d_j + c_j d_i)$
by a factor of $(m_\tau/m_\mu)^2$ -- the elements $\hat{Y}_{i3}$ are larger than $\hat{Y}_{i2}$ by
a factor or $(m_\tau/m_\mu)$.  So let us diagonalize $M_\nu$ dropping the subleading $(c_i d_j+c_jd_i)$
terms.  That is, we diagonalize the matrix $M_\nu^0$ with elements
\begin{equation}
(M_\nu^0)_{ij} = \kappa f_{23} (a_i b_j + a_j b_i)~.
\label{Mnu0}
\end{equation}
In addition to the trace being zero, $M_\nu^0$ has its (3,3) entry zero. Thus the the (1,1) and (2,2)
entries are equal and opposite.  Diagonalizing $M_\nu^0$ is achieved by the orthogonal transformation
\begin{equation}
U^T M_\nu^0 U = M_\nu^{\rm diag} = {\rm diag}.\left\{ \left(\frac{m_\tau^2\kappa f_{23}}{v}\right) \frac{\sqrt{\tan^2\beta + x^2 \cot^2\beta}}{1+x^2},~
-\left(\frac{m_\tau^2\kappa f_{23}}{v} \right) \frac{\sqrt{\tan^2\beta + x^2 \cot^2\beta}}{1+x^2},~0   \right\}
\label{eig}
\end{equation}
where the matrix $U \equiv U_{\rm PMNS}$, which is the PMNS matrix (up to signs), is found to be
\begin{eqnarray}
U_{\rm PMNS} = \left( \begin{matrix} \frac{1}{\sqrt{2}} (C_\chi C_\psi + S_\psi) & \frac{1}{\sqrt{2}} (C_\chi C_\psi - S_\psi) &
-S_\chi C_\psi \cr \frac{1}{\sqrt{2}} (C_\chi S_\psi - C_\psi) & \frac{1}{\sqrt{2}} (C_\chi S_\psi + C_\psi) & -S_\chi S_\psi \cr
\frac{S_\chi}{\sqrt{2}} & \frac{S_\chi}{\sqrt{2}} & C_\chi
\end{matrix}\right)~.
\label{PMNS}
\end{eqnarray}
Here $S_\chi = \sin\chi,\,C_\chi=\cos\chi$ and $S_\psi = \sin \psi,\, C_\psi = \cos\psi$, with
\begin{equation}
S_\psi = \frac{y}{\sqrt{1+x^2+y^2}},~~S_\chi = -\frac{(\tan\beta - x^2 \cot\beta)}{\sqrt{1+x^2}\sqrt{\tan^2\beta+x^2\cot^2\beta}}~.
\end{equation}
The matrix $U_{\rm PMNS}$ is a product of three rotation matrices obtained as follows.  The first rotation brings vector
$\vec{a}$ into the form $|\vec{a}|\, (0,\,1,\,0)^T$.  The transformed vector $\vec{b'} = (b_1',\,0,\,b_3')^T$ is then rotated to
the form $|\vec{b}|\, (1,\,0,\,0)^T$.  A third rotation by 45 degrees in the (1-2) sector brings the neutrino mass matrix $M_\nu^0$
to the diagonal form shown.

The crucial predictions of the model for the neutrino mixing parameters can now be stated:
\begin{itemize}
\item Neutrino mass hierarchy is inverted.
\item $\delta_{CP} = 0 ~{\rm or}~ \pi$.
\item  $|U_{\tau 1}| = |U_{\tau 2}|$.
\end{itemize}
Inverted mass hierarchy prediction follows from the two nearly degenerate mass eigenvalues.
$\delta_{CP}$ taking CP conserving values of 0 or $\pi$ follows from the reality of $M_\nu$ -- all phases could be absorbed into
fermion fields.  The equality of $|U_{\tau 1}|$ and  $|U_{\tau 2}|$ is evident in the form of $U_{\rm PMNS}$ shown in Eq. (\ref{PMNS}).
This can also be seen by the features noted on $M_\nu^0$, namely $(M_\nu^0)_{33} = 0$ and $(M_\nu^0)_{11}+ (M_\nu^0)_{22} = 0$, both
of which lead to this condition, once $m_1 = -m_2,\, m_3=0$ for the neutrino mass eigenvalues of Eq. (\ref{eig}) are used, along with
$\delta_{CP} = 0$ or $\pi$.

The leading two eigenvalues of $M_\nu$ are degenerate, but have opposite signs, as shown in Eq. (\ref{eig}).
When the subleading terms  $(c_i d_j + c_j d_i)$ terms in Eq. (\ref{Mnu}) in $M_\nu$ with relative
suppression factors of $(m_\mu/m_\tau)^2$ are included,  this degeneracy will
be lifted, and the solar mass-splitting of the right order will be induced, as we show below.  The effect of these subleading terms on
the PMNS matrix are tiny, so we should study first the consequences of the prediction $|U_{\tau 1}| = |U_{\tau 2}|$.  In the standard
parametrization of $U_{\rm PMNS}$, this prediction reads as
\begin{eqnarray}
s_{13} &=& t_{23} \frac{1-t_{12}}{1+t_{12}}, ~~{\rm or} ~~ s_{13} = -t_{23} \frac{1+t_{12}}{1-t_{12}} ~~~~~~(\delta_{CP} = \pi)~;\nonumber \\
s_{13} &=& t_{23} \frac{1+t_{12}}{1-t_{12}}, ~~{\rm or} ~~ s_{13} = -t_{23} \frac{1-t_{12}}{1+t_{12}} ~~~~~~(\delta_{CP} = 0)~.
\end{eqnarray}
Here $t_{23} = \tan\theta_{23}$, etc.  In the standard parametrization $\theta_{13}$ lies in the first quadrant, so the second solution in
each case above is inconsistent. Only the first solution with $\delta_{CP} = \pi$ will lead to
positive $s_{13}$ smaller than $t_{23}$.  Thus the model predicts $\delta_{CP} = \pi$.

\begin{figure}[h]
\centering
\includegraphics[scale=0.85]{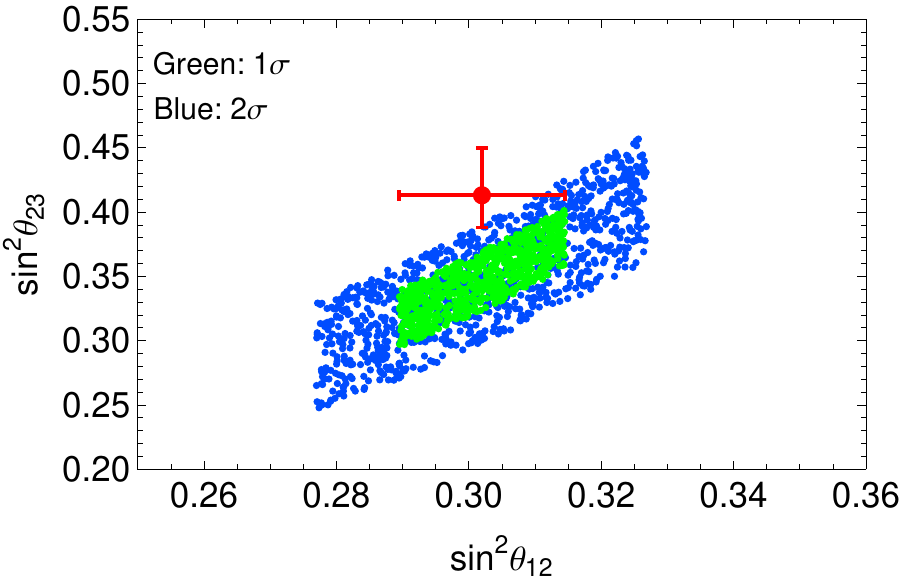}
\includegraphics[scale=0.85]{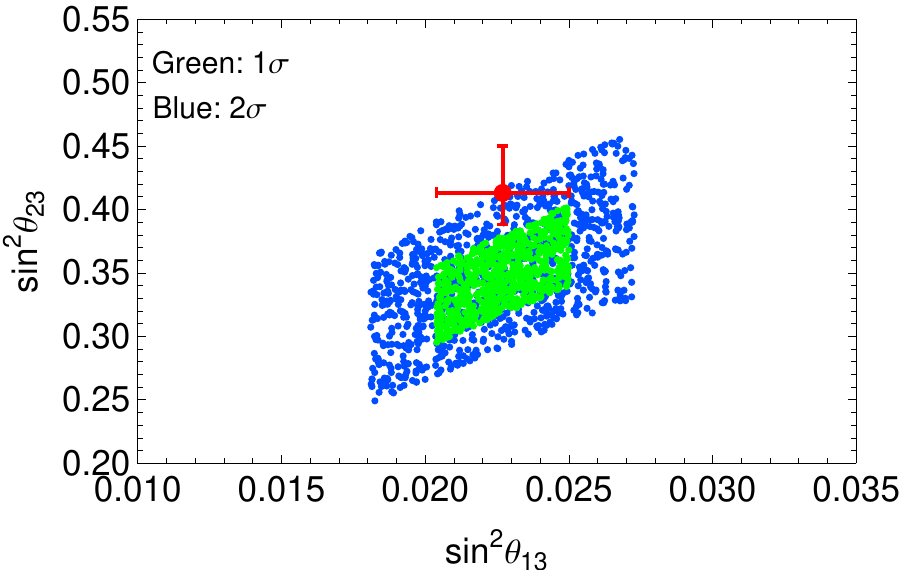}
\caption{Predicted value of $\sin^2\theta_{23}$ as functions of $\sin^2\theta_{12}$ (left panel) and $\sin^2\theta_{13}$ (right panel).
The green band shows the range of prediction with errors in the input quantities taken at 1 sigma,
while the blue band indicates the range at 2 sigma. The global fit to neutrino data is shown in red.}
\label{mixing}
\end{figure}

We plot the relation
\begin{equation}
s_{13} = t_{23} \frac{1-t_{12}}{1+t_{12}}
\label{relation}
\end{equation}
in Fig. (\ref{mixing}) in two planes, $\sin^2\theta_{23}$ versus $\sin^2\theta_{12}$, and
 $\sin^2\theta_{13}$ versus $\sin^2\theta_{23}$.  As inputs we use $\sin^2\theta_{12}$ and $\sin^2\theta_{13}$ obtained from a global fit of neutrino data \cite{GonzalezGarcia:2012sz, fogli}:
\begin{eqnarray}
\sin^2\theta_{12} = 0.302 \pm 0.0125, \quad \sin^2\theta_{13} = 0.0227 \pm 0.0023,
\end{eqnarray}
In Fig. (\ref{mixing}) we show the range of the predictions for the mixing angles in our model with one sigma and
two sigma error bars in the input quantities.    The best fit to the mixing angles from a global analysis of all neutrino data
is also shown in red along with its error bar.  The prediction of the model is found to be in very good agreement with data.
There is a preference for $\sin^2\theta_{12}$ to be slightly above the central value by about one $\sigma$.
Similarly, $\sin^2\theta_{23}$ is near 0.4, and cannot exceed about $0.45$ at two sigma.

Having established the consistency of the mixing angle prediction, we now turn to the subleading terms
of $M_\nu$ which is required to generate the solar mass splitting.  Treating the $(c_i d_j + c_j d_i)$ terms
as perturbations, we obtain
\begin{equation}
\Delta m^2_{\rm solar} = 4\left(\frac{m_\mu^2}{m_\tau^2}\right) \frac{xy \tan\beta \sqrt{\tan^2\beta + x^2 \cot^2\beta}}{(1+x^2+y^2)(\tan^4\beta+x^2)}\,
\left(\frac{m_\tau^2 \kappa f_{23}} {v}\right)^2~.
\label{solar}
\end{equation}
From Eq. (\ref{eig}) we also have
\begin{equation}
\Delta m^2_{\rm atm} \equiv m_3^2-m_1^2 = -\frac{\tan^2\beta+x^2\cot^2\beta}{(1+x^2)^2} \,\left(\frac{m_\tau^2 \kappa f_{23}} {v}\right)^2~.
\label{atm}
\end{equation}
The lightest neutrino mass $m_3$ is predicted in the model, which turns out to be tiny:
\begin{equation}
m_3 = \frac{1}{2}\, \frac{\Delta m^2_{\rm solar}} {|\Delta m^2_{\rm atm}|^{1/2}} \simeq 7.5 \times 10^{-4}~{\rm eV}.
\end{equation}
From these relations we can also compute the range for the effective mass parameter $m_{\beta \beta}$
for neutrinoless double beta decay. For this purpose we use the
atmospheric and solar mass splittings as input, obtained from the global fit:
\begin{eqnarray}
\Delta m_{\rm atm}^2 = (2.47 \pm 0.07) \times 10^{-3}~{\rm eV}^2, \quad
\Delta m_{\rm solar}^2 = (7.5 \pm 0.19) \times 10^{-5}~{\rm eV}^2.
\end{eqnarray}
By varying the input parameters within their 1 sigma range, we get
\begin{equation}
 m_{\beta\beta} \equiv  |\sum_{i=1-3}U_{ei}^2m_i | = (17.6-18.5) ~{\rm meV}.
\end{equation}
Here we also used the fact that the Majorana phases are zero, and that $m_1$ and $m_2$ have opposite CP parities.
We also demand that the value of $\sin^2\theta_{23}$ resulting from the model prediction is within 1 sigma of the best fit value.

The effective mass parameter $m_\beta$ that is measurable in beta decay end point spectrum, in experiments such as KATRIN,
is $m_\beta = \sum_i|U_{ei}|^2 m_i$, which is equal to $m_\beta = (1-|U_{e3}|^2)\, m_1 \simeq 0.049$ eV in our model.
Similarly, the sum of the neutrino masses, relevant for cosmology, is given by $m_{\rm cosmo} = m_1 + m_2 + m_3 \simeq 2\sqrt{|\Delta
m^2_{\rm atm}|} \simeq 0.1$ eV.

To determine the parameters of the model, we choose inputs values for $R = \Delta m^2_{\rm solar}/|\Delta m^2_{\rm atm}|$,
$|U_{e2}|$ and $|U_{e3}|$.  Using the relation for $R$ from Eqs. (\ref{solar})--(\ref{atm}), and the relations
$|U_{e2}|= |\frac{1}{\sqrt{2}} (C_\chi C_\psi - S_\psi)|$ and $|U_{e3}| = |-S_\chi C_\psi|$, for a given input choice we
solve for $(x,\,y,\,\tan\beta)$.  The third mixing angle $\theta_{23}$ is determined through Eq. (\ref{relation}).

Since the uncertainties in the input parameters are small, we can determine the model parameters $(x,\,y,\,\tan\beta)$
rather precisely.  With $|U_{e2}|^2 = 0.32$, $|U_{e3}|^2 = 0.0227$ and $R=1/32.9$, we find two separate solutions for the
parameters (up to signs):
\begin{eqnarray}
(i)&& (x,\,y,\,\tan\beta) = ( 0.038,\, 4.24,\, 0.189), \nonumber \\
(ii) && (x,\,y,\,\tan\beta) = (4.85,\, 21.0,\,  1.93)~.
\end{eqnarray}
These values can now be used to compute the Higgs Yukawa coupling matrix $\hat{Y}$, which would determine the structure
of flavor changing neutral currents.  For the two solutions we find this matrix to be
\begin{eqnarray}
(i)~~~~
\hat{Y} = \left(\begin{matrix} -2.95\times 10^{-7} & -0.00074 & -0.00049 \cr
-3.60 \times 10^{-6} & -0.003 & -0.002 \cr
-1.40 \times 10^{-7} & -0.0001 & 0.0018 \end{matrix} \right),
\label{Yhat1}
\end{eqnarray}
\begin{eqnarray}
(ii) ~~~~
\hat{Y} = \left(\begin{matrix} 5.64 \times 10^{-6} & -1.35 \times 10^{-5} & -0.0012 \cr
 -6.53 \times 10^{-8} &  0.001&  -0.005 \cr
 -3.26 \times 10^{-7} &  -0.0003 & -0.005
 \end{matrix} \right).
 \label{Yhat2}
\end{eqnarray}
With these coupling matrices, we can now determine FCNC rates, which we address in the next section.
We can also determine the overall coefficient of neutrino mass matrix from Eq. (\ref{atm}),
\begin{equation}
\kappa f_{23} = \{9.8 \times 10^{-9},~~2.1 \times 10^{-8}  \}~~ ({\rm for ~ cases}~ (i)~ {\rm and}~ (ii))~.
\end{equation}
The smallness of $\kappa f_{23}$ may be explained by choosing $\kappa$ and/or $f_{23}$ small.  A small
$\kappa$ is realized if the cubic scalar coupling coefficient $\mu$ in Eq. (\ref{V0}) is small, or if
the mass of one of the charged scalar $\eta^+$ or $H^+$ is  large.  As an illustration, choose
$\mu = 1$ GeV, $f_{23} = 0.01$, $m_{\eta^\pm} = 1.5$ TeV and $m_{H^\pm} = 500$ GeV.  This would
yield $\kappa f_{23} = 2.1 \times 10^{-8}$, consistent with solution $(ii)$.  Clearly, other choices are also
possible.  This shows that the smallness of neutrino masses can be explained in the present framework
without much tuning, even when the scale of new physics is near the TeV.

We note that the values of $(x,\,y,\,\tan\beta)$ determined via analytic approximation
can be used to solve the lepton mass and mixing problem exactly by numerical methods.
Excellent agreement is found for solution (i), and very good agreement is realized for
solution (ii) -- the difference in the two solutions being the largish $y$ in (ii).

\section {Lepton flavor violation mediated by Higgs bosons}

In our model both the Higgs doublets couple to lepton fields.  There are tree--level flavor changing
neutral currents mediated by the neutral Higgs bosons. The neutral Higgs and the charged Higgs can
also mediate lepton flavor violation through loop diagrams.  While the couplings of the charged Higgs $H^\pm$ and the
pseudoscalar Higgs $A^0$ to the leptons are uniquely fixed, couplings to the real scalar fields will involve an additional
mixing angle $\alpha$ defined through $H^0 = \cos\alpha \,{\rm Re}H_1^0 + \sin\alpha \,{\rm Re}H_2^0$, $h^0 = -\sin\alpha\,
{\rm Re} H_1^0 + \cos\alpha \,{\rm Re} H_2^0$.  For the special choice $\alpha = \beta - \pi/2$, the neutral field
$h^0$ will behave like the Standard Model Higgs field.  This choice of $\alpha$ is realized in the decoupling limit,
where the second Higgs doublet mass takes large values compared to $v$.  Perturbative realization of the decoupling
limit would prefer the presence of the soft $Z_4$ breaking term $m_{12}^2$ in Eq. (\ref{V0}).  In this limit, the
$A^0, \, H^0$ and $H^\pm$ fields will be nearly degenerate in mass.  In the unitary gauge, all components of $H_1$ and $H_2$
can be written in the decoupling limit as \cite{Branco:2011iw}
\begin{eqnarray}
{\rm Re}(H_1^0) = H^0 \sin\beta; \quad H_1^+ = H^+\sin\beta ; \quad {\rm Im}(H_1^0)=A^0 \sin\beta \\
{\rm Re}(H_2^0) = H^0 \cos\beta; \quad H_2^+ = -H^+\cos\beta ; \quad {\rm Im}(H_2^0)=-A^0 \cos\beta
\end{eqnarray}

The Yukawa couplings of the Higgs fields with the leptons in the decoupling limit (with $\alpha = \beta - \pi/2$ assumed
for $H^0$ coupling) is given by
\begin{equation}
{\cal L}_{\rm Yuk}^{(\ell)} = \frac{1}{\sqrt{2}} \ell_i \hat{Y}_{ij} \ell^c_j H^0 + \frac{i}{\sqrt{2}} \ell_i \hat{Y}_{ij} \ell^c_j A^0
+ \nu_{\ell_i} \hat{Y}_{ij} \ell^c_j H^- + \nu_{\ell_i} \hat{f}_{ij} \ell_j \eta^+  + h.c.
\label{Yukawa-1}
\end{equation}
where the elements of $\hat{Y}$ are determined as shown in Eqs. ({\ref{Yhat1})-(\ref{Yhat2}) from neutrino data.

From the structure of $\hat{Y}$ in Eqs. ({\ref{Yhat1})-(\ref{Yhat2}) it is clear that there will be lepton flavor violation mediated by
$A^0$ and $H^0$ scalars at the tree--level.  The process $\ell_i^- \to \ell_j^+ \ell_k^- \ell_l^-$ occurs at tree--level, as shown
in Fig. (\ref{tau3mu}) for the decay $\tau \rightarrow 3\mu$. Combining the contributions arising from $H^0$ and $A^0$ (see Eq. (\ref{Yukawa-1})),
with $m_{H^0} = m_{A^0}$ we obtain the rates for these processes to be
\begin{equation}
\Gamma (\ell_i \to 3\ell_j) = \frac{1}{64}
\frac{m_{\ell_i}^5}{192\pi^3}
\frac{|\hat{Y}_{ij}\hat{Y}_{jj}|^2+|\hat{Y}_{ji}\hat{Y}_{jj}|^2}{m_{A^0}^4}~.
\end{equation}

\begin{figure}[h]
\centering
\includegraphics[scale=0.8]{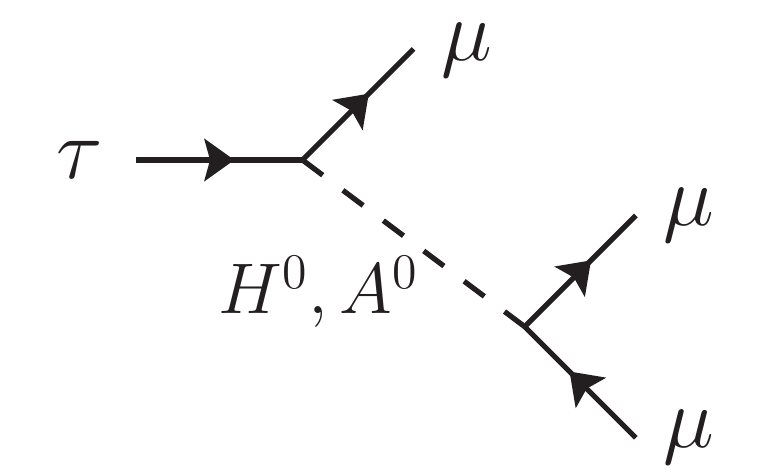}
\caption{Diagram leading to $\tau \to 3\mu$ decay by the exchange of neutral Higgs bosons.}
\label{tau3mu}
\end{figure}

Radiative decays $\ell_i \rightarrow \ell_j + \gamma$ arise in the model through one--loop diagrams mediated by the neutral Higgs
bosons $H^0$ and $A^0$ as well as the charged Higgs bosons $H^\pm$ and $\eta^\pm$.  Ignoring $H^\pm -\eta^\pm$ mixing, and setting
$m_{A^0} =m _{H^0} = m_{H^\pm}$ along with $\alpha = \beta - \pi/2$, the rate for these processes is given by
\begin{equation}
\Gamma(\ell_i \to \ell_j \gamma) = \frac{\alpha_{em} m_{\ell_i}^5}{(96\pi^2)^2}  \left[\frac{|\tfrac{1}{4}(\hat{Y}^T\hat{Y})_{ij}|^2}{m_{A^0}^4} + \frac{|\tfrac{1}{2}(\hat{Y}\hat{Y}^T)_{ij} - r^2(\hat{f}\hat{f}^T)_{ij}|^2} {m_{A^0}^4}\right],
\label{muegamma}
\end{equation}
where $r\equiv m_{A^0}^2/m_{\eta^+}^2$. The first term in Eq. (\ref{muegamma}) arises from the exchange of $H^\pm$ and $(A^0,H^0)$ with muon being right-handed, while the second term
is from the exchange of $(A^0,\, H^0)$ and $\eta^+$ with muon being left-handed. The relative minus sign is in the second set of terms is due to the fact that
the photon is emitted from the charged-lepton line in diagrams with the exchange of $(A^0, H^0)$,
while in $\eta^+$ exchange, it is emitted from the $\eta^+$ line.  Although it might appear that the two contributions interfere
destructively, with the values of $(x,\,y,\,\tan\beta)$ determined from neutrino oscillation data, it turns out that they interfere
constructively.  In our analysis we keep the contributions from  $(A^0, \,H^0,\, H^\pm)$ and not from $\eta^+$ since the coupling of $\eta^+$,
$f_{23}$, is not determined.  Our estimate will however be a lower limit on the rate for radiative decays.  Quite likely, the coupling
$f_{23}$ is small, or $\eta^+$ is heavy (see discussions after Eq. (\ref{Yhat2})), justifying our procedure.

We now examine the constraints arising from the processes $\tau \to 3\mu$ and $\mu \to e\gamma$. Their current limits are ${\rm BR}(\tau \to 3\mu)<2.1 \times 10^{-8}$ \cite{pdg} and ${\rm BR}(\mu \to e\gamma)<5.7 \times 10^{-13}$ \cite{Adam:2013mnn}. These limits translate into the constraints
\begin{eqnarray}
|\hat{Y}_{23}\hat{Y}_{22}|^2+|\hat{Y}_{32}\hat{Y}_{22}|^2 &<& 5.3 \times 10^{-7} \left(\frac{m_{A^0}}{150~{\rm GeV}}\right)^4, \nonumber \\
|\tfrac{1}{4}(\hat{Y}^T\hat{Y})_{21}|^2 + |\tfrac{1}{2}(\hat{Y}\hat{Y}^T)_{21} - r^2(\hat{f}\hat{f}^T)_{21}|^2 &<& 7.5 \times 10^{-10}\left(\frac{m_{A^0}}{150~{\rm GeV}}\right)^4.
\end{eqnarray}
These result suggest that for $m_{A^0} \simeq 150$ GeV, $\mu \to e\gamma$ branching ratio should be greater than  $6.6 \times 10^{-15}$, if we use
solution (ii) of Eq. (\ref{Yhat2}). This is consistent with present limits, and perhaps is within reach of MEG and other proposed
experiments. For solution (ii) the predicted branching ratio for $\tau \rightarrow 3\mu$ is (for $m_{A^0} = 150$ GeV) $1.2 \times 10^{-12}$.
For solution (i) of Eq. (\ref{Yhat1}), ${\rm BR}(\mu \rightarrow e\gamma) = 2.0 \times 10^{-15}$ and ${\rm BR}(\tau\rightarrow 3\mu) = 1.6 \times 10^{-12}$
corresponding to $m_{A^0} = 150$ GeV.  All other lepton flavor violation processes are much more suppressed.

\section{Higgs phenomenology}

With the knowledge of $\hat{Y}$ and $\tan\beta$, we can predict the branching ratios of $H^0,A^0,H^\pm$ into SM particles.
We start with Yukawa interactions in quark sector.
As noted in Sec. 2, a non-anomalous $Z_4$ symmetry suggests that the up-type and down-type quarks couple to the same Higgs doublet,
either $H_1$ or $H_2$. Let us first consider the case where the quarks couple to $H_1$.  In the decoupling limit ($\alpha = \beta - \pi/2$) the
Yukawa interactions of the quarks are given in the quark mass eigenbasis as
\begin{eqnarray}
{\cal L}_{\rm Yuk} &=& -\sum_{q} \frac{m_q}{v} \bar{q}qh^0 + \tan\beta \sum_{q} \left[ -\frac{m_q}{v} \bar{q}qH^0 + i\frac{m_q}{v} \bar{q}\gamma_5 q A^0 \right] \nonumber \\
&& - \tan\beta \frac{\sqrt{2}(V_{CKM})_{ij}}{v} \left[ (m_{u_i} \bar{u}_i P_L d_j - m_{d_j} \bar{u}_i P_R d_j)H^+  + h.c. \right]
\label{yukawa-q}
\end{eqnarray}
where $m_{u_i}$ and $m_{d_j}$ are up and down quark masses respectively and $P_{R,L}\equiv \tfrac{1}{2}(1\pm\gamma_5)$. It is important to notice that the interactions of $H^0$ and $A^0$ with quarks are flavor diagonal in our model.  When all the quarks couple to $H_2$, their couplings can be obtained from Eq. (\ref{yukawa-q}) by the replacement $\tan\beta \rightarrow \cot\beta,\,A^0 \rightarrow -A^0$ and $H^\pm \rightarrow -H^\pm$.

Among the two solutions obtained for $\tan\beta$, solution (i) ($\tan\beta = 0.19$) suggests that all quarks must couple to $H_1$. Otherwise
the top quark Yukawa coupling to $H^\pm$ would be of order 6 and non-perturbative.  Similarly, in solution (ii) ($\tan\beta = 1.9$), it is
preferable that all quarks couple to $H_2$, so that large top quark Yukawa coupling of order 2 is not generated.  We shall only consider
these two cases -- viz., solution (i) with all quarks coupling to $H_1$ and solution (ii) with all quarks coupling to $H_2$. We shall investigate
the Higgs boson branching ratios when the masses of $(H^0,\,A^0)$ are not too large, so that the decays $(H^0,\,A^0) \rightarrow t\bar{t}$ is not open. That is, we restrict this analysis to $(m_{A^0},\,m_{H^0},\,m_{H^\pm}) < 350$ GeV.

The mass of $H^\pm$ is constrained from the process $b \to s\gamma$. In type I two Higgs doublet model with $H_2$ coupling to up and down
quarks, $b\rightarrow s\gamma$ sets a constraint $\tan\beta > 1.8$ for  $m_{H^+}=300$ GeV \cite{Hermann:2012fc}.  For our solution (ii), this
requirement is satisfied with $\tan\beta = 1.9$. For our solution (i), since $H_1$ couples to all quarks, the constraint from $b \rightarrow
s\gamma$ is $\cot\beta > 1.8$ for $m_{H^\pm} = 300$ GeV.  This is also satisfied in our model, since we have $\cot\beta = 5.3$ in solution (i).

The partial decay rates for the Higgs boson decays are given by
\begin{align}
\Gamma(H^0,A^0 \to \bar{q}{q}) &= \frac{N_c m_q^2 \xi^2m_{A^0}}{16\pi v^2}(1-4m^2_q/m_{H^0}^2)^{3/2}; \quad {\Gamma}(H^0,A^0 \to \bar{\ell}_i\ell_j) = \frac{m_{A^0}}{16\pi}(\hat{Y}_{ij}^2 + \hat{Y}_{ji}^2); \nonumber \\
{\Gamma}(H^+ \to \bar{d}_ju_i) &= \frac{N_c\xi^2 m_{A^0}|(V_{CKM})_{ij}|^2(m_{u_i}^2+m_{d_j}^2) }{16\pi v^2}; \quad {\Gamma}(H^+ \to \ell^+_i\nu_{\ell_j}) = \frac{m_{A^0}}{16\pi}(\hat{Y}_{ji}^2),
\end{align}
with $N_c=3$ being the color factor and $\xi=\{\tan\beta,\,\cot\beta\}$ corresponding to solutions (i) and (ii).
For $m_{H^0}=m_{A^0}=300$ GeV, the dominant decay modes are $H^0,A^0 \to t\bar{t}^*,~b\bar{b},~\tau^+\tau^-,~\mu^+\mu^-,~\mu^\pm\tau^\mp$,
where the virtual $t^*$ decays as $t^* \to W^+b$.
Notice that there are no $H^0 \to W^+W^-,ZZ$ decays since the relevant couplings vanish in the limit of $\beta-\alpha=\pi/2$. In principle, $H^0$ could also decay into a pair of $h^0$. However, that coupling depends on a combination of quartic couplings which is unknown, and which
may be very suppressed.  We assume that this decay has a negligible rate.

We summarize the branching ratios of several decay channels below:

\noindent
{\bf Solution (i)}: $H_1$ couples to quarks:
\begin{align}
&{\rm BR}(H^0 \to t^*\bar{t})+{\rm BR}(H^0 \to t\bar{t}^*) = 0.15; \quad {\rm BR}(H^0 \to b\bar{b}) = 0.34; \quad {\rm BR}(H^0 \to \tau^+\tau^-) = 0.089; \nonumber \\
&{\rm BR}(H^0 \to \mu^+\mu^-) = 0.24; \quad {\rm BR}(H^0 \to \mu^+\tau^-) + {\rm BR}(H^0 \to \tau^+\mu^-) = 0.11;
\end{align}

\noindent
{\bf Solution (ii)}: $H_2$ couples to quarks:
\begin{align}
&{\rm BR}(H^0 \to t^*\bar{t})+{\rm BR}(H^0 \to t\bar{t}^*) = 0.22; \quad {\rm BR}(H^0 \to b\bar{b}) = 0.5; \quad {\rm BR}(H^0 \to \tau^+\tau^-) = 0.11; \nonumber \\
&{\rm BR}(H^0 \to \mu^+\mu^-) = 0.006; \quad {\rm BR}(H^0 \to \mu^+\tau^-) + {\rm BR}(H^0 \to \tau^+\mu^-) = 0.12;
\end{align}
In deriving these limits we use quark running masses at $\mu=M_t$ given in \cite{Babu:2009fd}.  We have also used the decay rate
for $A^0 \rightarrow t^* \bar{t}+ \bar{t}^* t = (1.97\,{\rm MeV})/\xi^2$ obtained from HDECAY \cite{Djouadi:1997yw}.
The charged-Higgs on the other hand, decays almost 100\%  of the time into $t\bar{b}$ in both solutions (i) and (ii).

We see that the branching ratios into leptons, especially into muons, is significant.  If the neutral Higgs particles are lighter than 300 GeV, their leptonic branching ratios may be even larger (for $\alpha = \beta - \pi/2$).  This will open up the discovery potential of such particles.
Higgs discovery with the prescribed properties can thus lend support to our model.  It should be noted that in the absence of soft $Z_4$ symmetry breaking, the second neutral Higgs boson cannot be much heavier than about 150 GeV. This is because both neutral scalar bosons have masses
of order $2 \lambda_1 v_1^2$ and $2 \lambda_2 v_2^2$ along the diagonal in the $2 \times 2$ mixing matrix.  In solution (i) we have $v_2 = 32$ GeV,
while in solution (ii) we have $v_1= 81$ GeV.  If the quartic scalar couplings are not much larger than one, the neutral scalars should be
relatively light, in the case of exact $Z_4$ symmetry.  With soft breaking of $Z_4$ this conclusion will not apply.

\section{Conclusions}

We have presented in this paper a simple model of radiative neutrino masses.  The model is a special case of the general Zee model.
We employed a family-dependent $Z_4$ symmetry that resulted in a total of four real parameters explaining the entire neutrino oscillation
data.  There are a variety of predictions in the neutrino sector.  The CP violating parameter $\delta_{CP}$ is predicted to be $\pi$.  Most interestingly, one of the neutrino oscillation angles is determined in terms of the other two angles.  This nontrivial relation is found to be consistent with current data.  Future precision determinations of $\sin^2\theta_{23}$ and $\sin^2\theta_{12}$  could serve as a test of the model.
The model prefers $\sin^2\theta_{23} \simeq 0.4$ and not more than $0.45$ at 90\% CL.  There is slight preference for $\sin^2\theta_{12}$
to be above the current central value by about one sigma.

The model employs two Higgs doublets and a charged singlet.  A crucial parameter that enters in two Higgs doublet models is the VEV ratio
$\tan\beta$.  We are able to determine its value from neutrino oscillations. We found that $\tan\beta = 0.19$ or $1.9$.  The branching
ratios of the neutral Higgs bosons of the model into fermions are then completely determined.  We found that leptonic decays, involving the muon,
can be significant, which can potentially raise the reach for such particles at the LHC.  The charged and neutral Higgs bosons also mediate
leptonic flavor violation, with $\mu \rightarrow e\gamma$ possibly within reach of proposed experiments.  The decay $\tau \rightarrow
3\mu$, which arise at the tree-level is also significant.  Lepton flavor violation with prescribed branching ratios would be yet
another test of the model.

\section*{Acknowledgements}
The work of KSB is supported in part by the US Department of Energy Grant No. DE-FG02-04ER41036. The work of JJ is supported by the Slovenian Research Agency.


\begin{thebibliography}{99}

\bibitem{Zee:1980ai}
  A.~Zee,
  Phys.\ Lett.\ B {\bf 93}, 389 (1980)
  [Erratum-ibid.\ B {\bf 95}, 461 (1980)].


 \bibitem{Wolfenstein:1980sy}
  L.~Wolfenstein,
  Nucl.\ Phys.\ B {\bf 175}, 93 (1980).

  \bibitem{Glashow:1976nt}
  S.~L.~Glashow and S.~Weinberg,
   Phys.\ Rev.\ D {\bf 15}, 1958 (1977).
   
   \bibitem{frampton}
   See for e.g:
    A.~Y.~.Smirnov and M.~Tanimoto,
  Phys.\ Rev.\ D {\bf 55}, 1665 (1997);
   C.~Jarlskog, M.~Matsuda, S.~Skadhauge and M.~Tanimoto,
  Phys.\ Lett.\ B {\bf 449}, 240 (1999);
   P.~H.~Frampton and S.~L.~Glashow,
  Phys.\ Lett.\ B {\bf 461}, 95 (1999).




   \bibitem{He:2003ih}
   Y.~Koide,
  Phys.\ Rev.\ D {\bf 64}, 077301 (2001);
  X.~-G.~He,
  Eur.\ Phys.\ J.\ C {\bf 34}, 371 (2004).




\bibitem{alternative1}
  A.~Zee,
  Nucl.\ Phys.\ B {\bf 264}, 99 (1986);
  K.~S.~Babu,
  Phys.\ Lett.\ B {\bf 203}, 132 (1988).

\bibitem{alternative2}
K.~S.~Babu and C.~Macesanu,
  Phys.\ Rev.\ D {\bf 67}, 073010 (2003); M.~Nebot, J.~F.~Oliver, D.~Palao and A.~Santamaria,
  Phys.\ Rev.\ D {\bf 77}, 093013 (2008);
   D.~Aristizabal Sierra and M.~Hirsch,
  JHEP {\bf 0612}, 052 (2006);
   E.~Ma,
  Phys.\ Rev.\ D {\bf 73}, 077301 (2006);
   E.~Ma,
  Phys.\ Rev.\ D {\bf 73}, 077301 (2006);
   M.~Aoki, S.~Kanemura and O.~Seto,
  Phys.\ Rev.\ Lett.\  {\bf 102}, 051805 (2009);
    K.~S.~Babu and J.~Julio,
  Nucl.\ Phys.\ B {\bf 841}, 130 (2010);
   F.~Bonnet, M.~Hirsch, T.~Ota and W.~Winter,
  JHEP {\bf 1207}, 153 (2012);
  P.~W.~Angel, N.~L.~Rodd and R.~R.~Volkas,
  Phys.\ Rev.\ D {\bf 87}, 073007 (2013);
   P.~W.~Angel, Y.~Cai, N.~L.~Rodd, M.~A.~Schmidt and R.~R.~Volkas,
  arXiv:1308.0463 [hep-ph].


\bibitem{discrete}
 L.~M.~Krauss and F.~Wilczek,
  Phys.\ Rev.\ Lett.\  {\bf 62}, 1221 (1989).


\bibitem{traceless}
X.~-G.~He and A.~Zee,
  Phys.\ Rev.\ D {\bf 68}, 037302 (2003);
W.~Rodejohann,
  Phys.\ Lett.\ B {\bf 579}, 127 (2004); B.~Brahmachari and S.~Choubey,
  Phys.\ Lett.\ B {\bf 642}, 495 (2006).


\bibitem{GonzalezGarcia:2012sz}
  M.~C.~Gonzalez-Garcia, M.~Maltoni, J.~Salvado and T.~Schwetz,
  JHEP {\bf 1212}, 123 (2012).

  \bibitem{fogli}
   G.~L.~Fogli, E.~Lisi, A.~Marrone, D.~Montanino, A.~Palazzo and A.~M.~Rotunno,
  Phys.\ Rev.\ D {\bf 86}, 013012 (2012).

\bibitem{Branco:2011iw}
J.~F.~Gunion and H.~E.~Haber,
  Phys.\ Rev.\ D {\bf 67}, 075019 (2003);
  For a review of general two-Higgs-doublet models see:
  G.~C.~Branco, P.~M.~Ferreira, L.~Lavoura, M.~N.~Rebelo, M.~Sher and J.~P.~Silva,
  Phys.\ Rept.\  {\bf 516}, 1 (2012).



\bibitem{pdg}
J.~Beringer {\it et al.}  [Particle Data Group Collaboration],
  Phys.\ Rev.\ D {\bf 86}, 010001 (2012).

\bibitem{Adam:2013mnn}
  J.~Adam {\it et al.}  [MEG Collaboration],
  arXiv:1303.0754 [hep-ex].

  \bibitem{Hermann:2012fc}
  T.~Hermann, M.~Misiak and M.~Steinhauser,
  JHEP {\bf 1211}, 036 (2012).


\bibitem{Babu:2009fd}
 Z.~-z.~Xing, H.~Zhang and S.~Zhou,
  Phys.\ Rev.\ D {\bf 77}, 113016 (2008);
K.~S.~Babu,
  arXiv:0910.2948 [hep-ph].


\bibitem{Djouadi:1997yw}
  A.~Djouadi, J.~Kalinowski and M.~Spira,
  Comput.\ Phys.\ Commun.\  {\bf 108}, 56 (1998).

\end{thebibliography}
\end{document}